\documentclass[twocolumn,showpacs,preprintnumbers,amsmath,amssymb,aps,prb]{revtex4}
\usepackage{graphicx}
\begin{document}

\title{Defect Fluctuations and Lifetimes in Disordered Yukawa Systems 
} 
\author{C. Reichhardt and C.J. Olson Reichhardt} 
\affiliation{ 
Theoretical Division and Center for Nonlinear Studies,
Los Alamos National Laboratory, Los Alamos, New Mexico 87545}

\date{\today}
\begin{abstract}

We examine the time dependent defect fluctuations and lifetimes for a 
bidisperse disordered assembly of Yukawa particles.  
At high temperatures, the noise spectrum of fluctuations 
is white and the coordination number
lifetimes have a stretched exponential distribution.
At lower temperatures, the system dynamically freezes,
the defect fluctuations exhibit a $1/f$ spectrum, and there is 
a power law distribution of the coordination number 
lifetimes. 
Our results indicate that topological defect fluctuations may be
a useful way to characterize disordered systems. 
\end{abstract}
\maketitle

\vskip2pc

\section{Introduction}
 
In two dimensional systems of interacting repulsive particles such
as dusty plasmas \cite{Lin,Woon,Liu} 
and colloidal assemblies \cite{Rice,Dullens,Levy,Reichhardt}, 
heterogeneous particle motions have been observed
in the
dense liquid phase.
The heterogeneity appears
close to the disordering transition where the system changes
from having predominantly hexagonal ordering  
to being
heavily defected. Recently, it was demonstrated that in the disordered 
regime where dynamical heterogeneities are present,
the time dependent density fluctuations of  
the topological defects exhibits a $1/f$ power spectrum \cite{Reichhardt}. 
The heterogeneous motion disappears
at higher temperatures, 
and simultaneously the noise spectrum changes to a white
form and the noise power
is strongly reduced, 
indicating a lack of correlations in the fluctuations.
It was argued that motion in the regime with 
dynamical heterogeneities occurs in a correlated manner
and is spatially concentrated in areas containing a higher density of
topological defects. 
As a result, the creation and annihilation of
topological defects is strongly correlated,
leading to the $1/f$ noise signature. Recent experiments 
on two-dimensional colloidal systems 
exhibiting heterogeneous motion have confirmed that the
colloids with more rapidly changing
coordination number are correlated with the regions of motion \cite{Dullens}.  

In the systems mentioned so far, the ground state at  
low temperatures is an ordered hexagonal lattice; however, 
it is known that
intrinsically disordered or 
glassy systems also exhibit dynamical heterogeneities \cite{Kob,Weeks}. 
A natural question to ask
is how do topological defect fluctuations behave in systems that  
are inherently disordered and that have no transition to an 
ordered state as the temperature is lowered.
Although there is no ordering transition in an intrinsically disordered system,
there can be a temperature at which a dynamical slowing down occurs,
giving rise to glassy type behaviors. 
It would be interesting to study how or if the defect fluctuations change 
as dynamical freezing  
is approached. In the disordered system,
the length scale
of the dynamical heterogeneities 
increases as the temperature is lowered \cite{Kob,Weeks}. If the 
moving regions
are associated with more highly defected regions, this may also be
reflected in the defect fluctuations.    

In this work we examine a two-dimensional system of a bidisperse mixture of
Yukawa particles. 
Previously, this system has been shown to form a disordered state at
all temperatures \cite{Hastings}. We analyze the time dependent defect 
fluctuations,
noise spectra, noise power, 
and the coordination number lifetimes for varied temperatures. At high 
temperatures where there is no
heterogeneous motion, 
the defect fluctuations have a white power spectrum, the coordination
number lifetimes are very short, and there is an exponential decay
in the distribution of lifetimes.  As the temperature is lowered,
the defect fluctuations show 
a $1/f$ power spectrum with increased noise power.
We also find a peak in the noise power at a finite
temperature we label $T_{n}$. Near 
this temperature, the motion is highly heterogeneous and
the coordination lifetimes are power
law distributed. 
For temperatures below $T_{n}$, the noise power is 
reduced and the coordination number lifetimes become
extremely long.
We also probe the system by examining the motion 
of a single particle driven by an external force. At $T = 0$ 
there is a well defined
threshold force for motion \cite{Hastings}.
At finite temperatures, creep occurs; however, there is
a well defined kink in the velocity force curves at a threshold force
which vanishes at $T_{n}$. 
These results suggest that measuring topological
defect fluctuations may be a 
useful probe for understanding glassy and jamming behaviors in disordered
systems. 

\section{Simulation}

We consider a two-dimensional system with sides of length $L$ and with
periodic boundary conditions in the $x$ and $y$ directions. 
The system contains $N$ particles interacting via a Yukawa or screened 
Coulomb potential. For any two particles $i$ and $j$ of charge
$q_i$ and $q_j$ located at 
positions ${\bf r}_{i}$ and ${\bf r}_{j}$, the pair interaction potential is 
$ V(r_{ij}) = q_{i}q_{j} \exp(-\kappa r_{ij})/r_{ij}$.
Here $r_{ij}=|{\bf r}_{i} - {\bf r}_{j}|$ 
and
$1/\kappa$ is the screening length which is set equal to $2.0$. 
We consider a 50:50 mixture of particles with charges
$q_1$ and $q_2$ where
$q_{1}/q_{2} = 1/2$. We have previously shown that 
this system forms a disordered
assembly at all temperatures \cite{Hastings}.
The particles obey overdamped dynamics and the 
time dependent particle positions and velocities 
are obtained by integrating the overdamped 
Brownian dynamics \cite{brownian} equations of motion, which for 
a single particle $i$ is 
\begin{equation}
\eta \frac{d {\bf r}_{i}}{dt} = {\bf F}_{i} +  {\bf F}_{T} + {\bf F}_d
\end{equation}
Here
$\eta$ is the damping coefficient and
${\bf F}_{i} = -\sum_{j \neq i}^{N}\nabla_i V(r_{ij})$.  
The particle-particle interactions are cut off 
at lengths larger than $2/\kappa$ and are calculated using
a neighbor index method
for computational efficiency. 
Further cutoff lengths have negligible effects. 
The thermal force is modeled as random Langevin kicks 
with $\langle{\bf F}^{T}_{i}\rangle = 0$ and
$\langle{\bf F}^{T}(t){\bf F}^{T}(t^{\prime})\rangle = 2\eta k_{B}T\delta(t - t^{\prime})$. 
The initial conditions are obtained by simulated annealing.
For our parameters, if all
the charges are set equal to the larger charge $q_2$, 
melting as defined by the onset of defects 
occurs at a temperature of $T = 4.0$. 
The driving force ${\bf F}_d$ is applied to a single particle only, such
that ${\bf F}_d=f_d{\bf {\hat x}}$ for the driven particle and ${\bf F}_d=0$
for all other particles.  For most of this work, we set $f_d=0$.
We anneal from a temperature of $T = 8.0$ down 
to a final temperature in steps of $\Delta T=0.25$ 
and sit at each increment temperature for over $10^6$ 
Brownian dynamics (BD) time steps. 
After annealing, we analyze the 
topological defect density and particle motions at a fixed temperature.  
Unless otherwise noted, we consider systems of size $L=32$ containing
$N=830$ particles.

\begin{figure}
\includegraphics[width=3.5in]{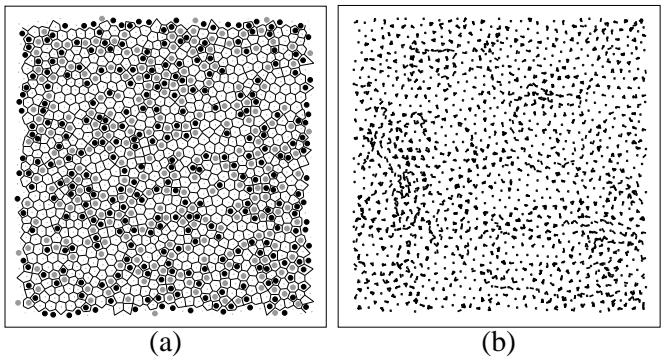}
\caption{
(a) Voronoi construction for a snapshot of a bidisperse
particle system at $T = 1.5$.  White polygons have 6 neighbors, while
black and gray polygons have 5 and 7 neighbors respectively.
(b) The particle positions (black dots) and trajectories (black lines) for
the same system during $10^5$ simulation time steps.    
}
\end{figure}

In Fig.~1(a) we show the topological defects as obtained from a Voronoi 
construction for a
system at $T = 1.5$.
The Voronoi construction forms a polygon around each particle
and is used to define the coordination number $C_n$ of the particle. 
Particles with five, six, and seven neighbors 
corresponding to $C_n=5$, 6, and 7 are marked black, white, and
gray, respectively.
The system is also highly defected at other 
temperatures and there is little change in the
average defect density with temperature; however, the time dependent 
fluctuation rate of the defects changes significantly 
with temperature as we show later.  
In Fig.~1(b) we illustrate the particle trajectories during
$10^5$ BD time steps at $T=1.5$.  The motion is highly heterogeneous, and
certain regions show motion during this time interval while other 
regions do not. 
At higher temperatures the
motion becomes more homogeneous throughout the sample. 

\begin{figure}
\includegraphics[width=3.5in]{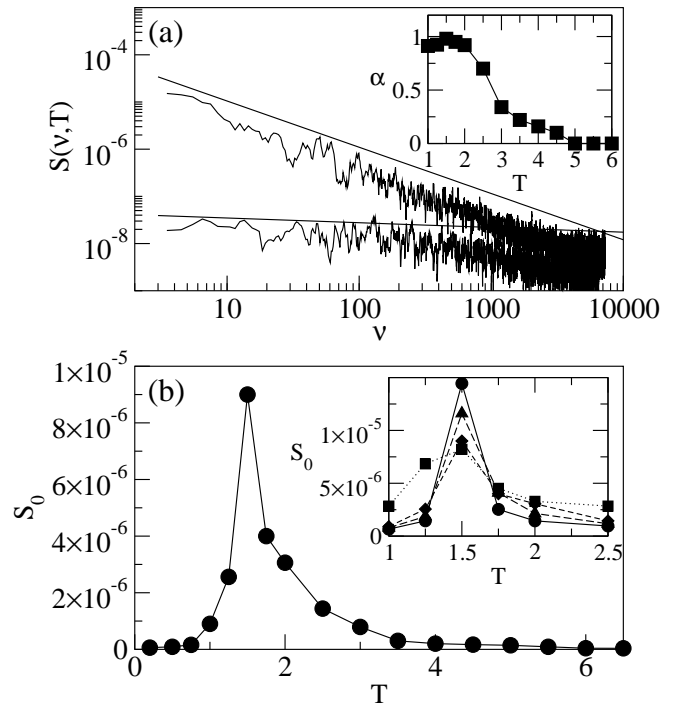}
\caption{
(a) Power spectra $S(\nu,T)$ obtained from the time series of the 
topological defect density fluctuations  
$P_6(t)$ obtained at $T = 1.5$ (upper curve)
with a fit (solid line) of $\alpha = 0.98$ 
and $T = 4.5$ (lower curve) 
with a fit (solid line) of $\alpha = 0.1$. The $T = 4.5$ curve has been shifted
down for presentation purposes.  Inset: the fitted values 
of $\alpha$ obtained from the power spectra 
vs $T$. (b) The noise power $S_{0}$ vs $T$. 
Inset: $S_{0}$ for $T = 1.0$ to $T = 2.5$ for
varied system sizes $L = 16$ 
(squares),
$32$ (diamonds), $48$ (triangles), 
and $54$ (circles). 
}
\end{figure}

\section{Defect Fluctuations and Power Spectra} 

In order to 
characterize the defect       
fluctuations, we perform a series of simulations at different 
$T$ and examine the
time dependent density of six-fold coordinated particles, $P_{6}(t)$,
defined as $P_6(t)=N^{-1}\sum_{i=1}^{N}\delta(C_n^i(t)-6)$,
where $C_n^i$ is the coordination number of particle $i$. 
The power spectrum of the resulting time series is defined as
\begin{equation} 
S(\nu,T) = \left|\int P_{6}(t)  e^{-2\pi i\nu t}dt\right|^2
\end{equation}
The noise power $S_{0}$ is defined as the average value of the noise spectrum 
at a particular value of the frequency $\nu$.
After annealing we wait $10^4$ time steps before acquiring the time
series $P_6(t)$ to avoid any transient behaviors.  
In Fig.~2(a) we show the power spectra of $P_6(t)$
for $T = 1.5$ and $T = 4.5$.
At $T = 1.5$  the power spectrum has a $1/f^{\alpha}$ form with
$\alpha \approx 0.98$, as indicated by the upper solid line.
For $T = 4.5$ the noise spectrum is approximately white and
$\alpha \approx 0.1$. 
We plot the evolution of $\alpha$ with $T$ in the inset of Fig.~2(a).
For $T \leq 2.0$, we find $1/f$ noise with $\alpha \approx 1$, while
at high $T$, the noise spectrum becomes white with $\alpha \approx 0$.
The particle trajectories also 
indicate that the motion is no longer heterogeneous for $T > 2.0$.  
In monodisperse systems near the disorder transition, 
a similar $1/f$ noise signal was observed when the particle motions
were highly heterogeneous \cite{Reichhardt}. 
We note that for $T < 1.5$ 
there is very little particle motion and the 
defect fluctuations
are strongly suppressed. This can be more clearly 
seen by examining the noise power $S_{0}$
at fixed $\nu=5$ as shown in Fig.~2(b). At high temperatures, $S_{0}$ is low,
but it increases rapidly to a peak value as the temperature is reduced  
to $T = T_n = 1.5$, and then decreases quickly for $T < 1.5$.
We associate the peak in $S_0$ at $T_n$ with the temperature at which the
system begins to undergo a dynamical freezing.  

In the inset of Fig.~2(b), we plot the variation in $S_0$ for systems of
different sizes ranging from $L=16$ to 54.
As $L$ increases, the peak in the noise power becomes more pronounced.
Additionally, although the height of the peak 
in $S_0$ increases with $L$, at higher or lower $T$ the noise
power at the non-peak temperatures 
is reduced with increasing $L$.  
For temperatures away from $T = 1.5$, the 
individual power spectrum curves have a rollover
to a white spectrum at the lowest frequencies.
At $T = 1.5$, the $\alpha = 1.0$ behavior persists down to
the lowest frequencies, as shown in Fig.~2(a). 
The suppression of the noise power away from $T=1.5$ in the
larger systems can be understood as arising from the fact that there is a
length scale $l$ associated with the correlated creation or annihilation of
defects.  The fluctuations with the largest $l$ 
correspond to the lowest frequencies.
If $l$ is constant, then when the
system size is small or comparable to $l$, 
the fluctuations will have a $1/f$ character
down to the lowest observable frequencies and  
the noise power will be increased away from the peak in $S_0$, 
as seen for $L = 16$ in the inset of Fig.~2(b).
As the system is made larger, the fluctuations begin
to average out, a low frequency cutoff appears
in the $1/f$ spectrum, and the noise power is reduced. 
Near the dynamical freezing temperature $T_n$, 
the fluctuations grow with the system size and  
$1/f$ noise is present for all arbitrarily large $L$ and low $\nu$.  
For $T < T_{n}$, the fluctuations are again cut off 
and the noise power is reduced.   
We have compared our results against simulations of 
collections of non-interacting particles, where the
motion is homogeneous and 
diffusive at all temperatures. 
In this case, there is no peak in $S_0$ and the
spectra are white for all temperatures. 

\begin{figure}
\includegraphics[width=3.5in]{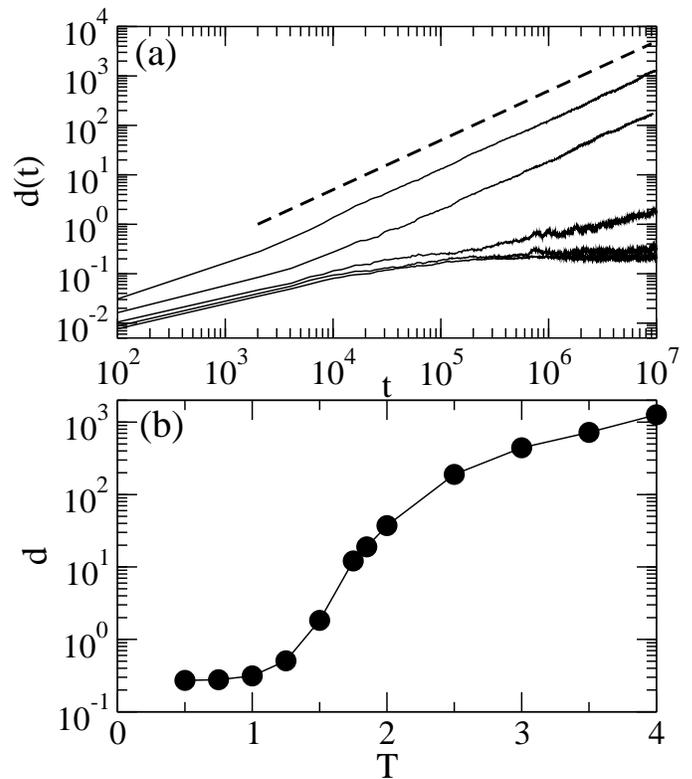}
\caption{ 
(a) 
The measure of the particle displacements $d(t)$ 
vs $t$  
for 
(top curve to bottom curve) $T = 3$, 2.5, 1.5, 1.0, and 0.5.
The dashed line is a linear fit showing the behavior expected for 
a system undergoing normal diffusion.      
(b) The value of $d(t)$ at $t = 10^{7}$ time steps 
measured for the same system as a function of $T$.
}
\end{figure}
 
\section{Diffusion} 

To quantify the amount of motion present in the system as a function
of temperature, we measure the diffusion of the particles by computing
the square displacements of the particles from an initial
position as a function of time,  
\begin{equation}
d(t) = \sum^{N}_{i=1}|{\bf \tilde{r}}_{i}(t) - {\bf \tilde{r}}_{i}(t_{0})|^2 ,
\end{equation}
where ${\bf \tilde{r}}_i(t)$ is the position of particle $i$ at time $t$
with the effect of crossing the periodic boundary conditions removed.
For particles undergoing normal 
diffusion, $d(t) \propto t$. In Fig.~3(a) we plot $d(t)$ 
at $T = 3.0$, 2.5, 1.5, 1.0, and $0.5$ 
from top to bottom. The data was taken over
a period of $10^7$ time steps which started after the system had entered
a stationary state. 
In general, we find a transient time period which lasts for
$10^3$ time steps after annealing.  This transient behavior was excluded
from the analysis so that $t_0>10^3$ time steps.
For $T > 1.5$,  the long time behavior of the system is consistent with
normal diffusion, as indicated by the dashed line
showing a linear slope in Fig.~3(a). For $T < 1.5$, 
within our time frame $d(t)$ saturates to a constant value, 
while at $T = 1.5$ the long time behavior is more consistent with
subdiffusion,
$d(t) \propto t^{\alpha}$ where $\alpha < 1.0$. We note that this
agrees with the measures of the noise fluctuations which indicate that
the $1/f$ noise which appears at $T =  T_n = 1.5$ 
is associated with the presence 
of system-wide slow dynamics. 
For $ T < 1.5$ the fluctuations are instead associated with the rattling
motions of single particles rather than long time system wide motions.        
In Fig.~3(b) we plot the value of $d$ obtained at the end of $10^7$ steps.
For $T < 1.5$, $d$ saturates to a constant value
since within our time frame the particles are no longer diffusing 
significantly, while $d$ increases with $T$ for higher $T$.  

We note that since we are restricted to performing simulations over
finite periods of time, we cannot be certain that the subdiffusive motion at
$T=1.5$ does not cross over to regular diffusion at extremely long
times.  If this were the case, the power spectrum of the corresponding
very long time series would show a low frequency cutoff in the $1/f$
signature at the time scale where the regular diffusion begins to 
dominate.  What we can conclude from our data is that over the time regimes
which we are able to access, the appearance of the 
$1/f$ defect fluctuations is correlated with the subdiffusive behavior.   

\begin{figure}
\includegraphics[width=3.5in]{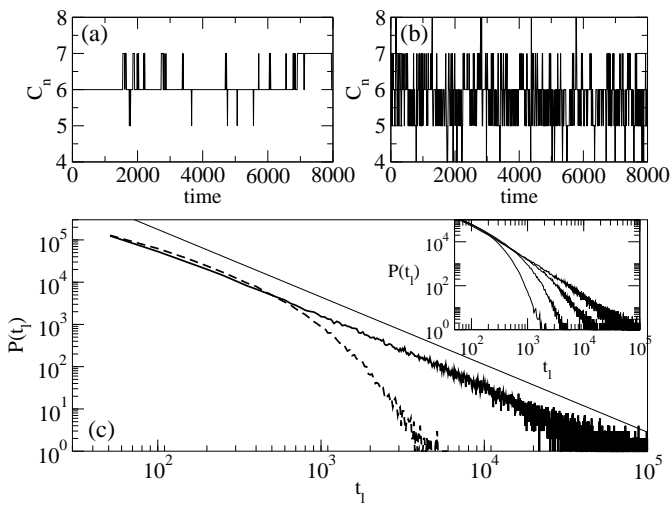}
\caption{ 
(a) The coordination number $C_n$ vs 
time for a single particle at $T = 1.5$ showing occasional changes. (b) 
$C_n$ vs time for a single particle at $T = 3.5$ showing much more 
rapid changes.  
(c) The histogram of the coordination number lifetimes 
$P(t_l)$
for $T = 3.5$ (lower curve) and $T = 1.5$ (upper curve).
The solid line is a
power law fit with $\beta = 1.6$. Inset: 
$P(t_l)$ at $T = 5.0$, 3.5, 2.5, and $1.5$, from left to right.
}
\end{figure}

\section{Defect Lifetimes} 

We next consider the lifetimes of the coordination numbers.
In Fig.~4(a) we show the coordination number
$C_{n}$ as a function of time for one particle
in a system at $T = 1.5$. 
During this time frame, the particle is predominantly sixfold coordinated
with $C_n=6$, while there are occasional jumps to a configuration with
$C_n=5$ or 7.
In some cases, such as for $t > 7000$, the particle maintains $C_n \ne 6$
for a longer period of time.
We obtain a similar plot regardless of which particle in the system is 
selected.
In Fig.~4(b) we plot $C_{n}$ vs time for    
the same particle at $T = 3.5$, showing that 
the particle coordination number is changing much more rapidly. 
The coordination lifetimes $t_l$
are obtained by measuring the length of time that elapses
between changes in $C_n$ for a given particle.
We histogram this quantity over all the particles, and plot the resulting
$P(t_l)$ 
in Fig.~4(c) 
for two systems at $T = 3.5$ (lower curve) and $T = 1.5$ (upper curve). 
The straight solid line is a power law fit to
$P(t_l)\propto t_l^{-\beta}$ 
with $\beta = 1.6$, 
while the curve at $T = 3.5$ is fit well to a stretched exponential.  
In the inset of Fig.~4(c) we 
plot $P(t_l)$ for $T = 5.0$, 3.5, 2.5, and $1.5$, 
which shows that for the higher temperatures
the average lifetime $\langle t_l\rangle$ is shorter. 
For $T = 5.0$, $P(t_l)$ can be fit reasonably well to an
exponential distribution.
For the intermediate temperatures, we can fit the data with a stretched
exponential form, and at $T = 1.5$ we obtain a power law fit.
For $ T < 1.5$, the lifetimes become extremely long 
and we observe close to a bimodal distribution where there  
are a few particles jumping back and forth giving 
rise to short values of $t_l$,
while $C_n$ of the other particles does not change for the duration of 
the simulations over times of $2.5\times 10^7$ BD steps. 

\begin{figure}
\includegraphics[width=3.5in]{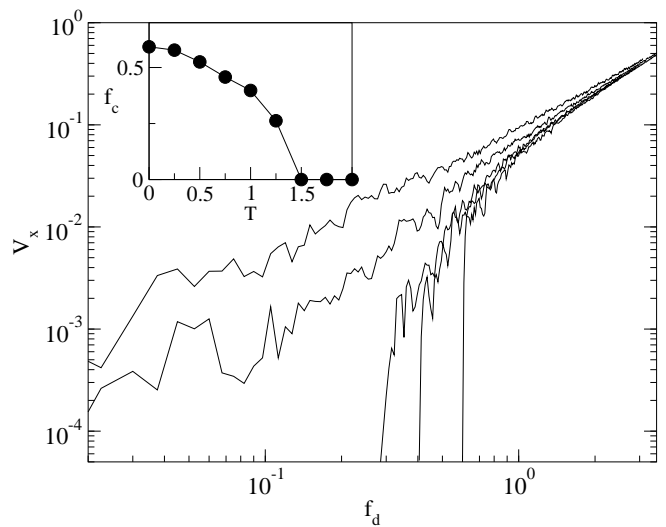}
\caption{
The velocity $V_{x}$ vs $f_{d}$ for a single driven probe particle for 
$T = 3.5$, 2.0, 1.25, 1.0, and $0$ from left to right.
The peak
in the noise power of the defect fluctuation power spectra occurs
at $T = T_n = 1.5$, which also corresponds to the onset of the subdiffusive
motion. 
Inset: the 
threshold force $f_{c}$ extracted from the velocity-force curves
vs $T$.    
}
\end{figure}

\section{Yielding Behavior} 

The results from Figs.~2, 3 and 4 suggest that near $T = 1.5$, some type of 
dynamical crossover or freezing occurs.
We can also probe the system by slowly driving a single particle 
through it.
If the system acts rigidly at low drives, then there is a threshold force
$f_{c}$ above which the driven particle moves with respect to the background. 
Previous work on this system at $T = 0$ has characterized this threshold force
\cite{Hastings}. 
Experiments have also shown evidence for a threshold force to motion for 
disordered colloidal systems \cite{Habdas}.  
In general, we find that creep can occur at low but finite temperatures;
however, a threshold force can still be identified by a sudden increase or
kink in the particle velocity vs applied
force. In Fig.~5 we plot the velocity $V_x$ vs applied drive $f_{d}$ for
a single probe particle which has a charge $q/q_{1} = 5.0$ moving through
a disordered assembly of particles 
at temperatures ranging from $T=0$ to $T=3.5$. 
If the surrounding particles are absent, 
the probe particle moves linearly with $V \propto F_{d}$. 
The velocity-force curves are obtained by slowly increasing the applied 
drive $f_d$. 
For $T < 1.5$, $V_{x}$ shows a sharp downward concavity at 
finite $f_{d}$. 
We define the threshold force $f_{c}$ as the value at which  
$V_{x}=5\times 10^{-5}$.
For $ T > 1.5$, the curves show a slight positive concavity and 
if we linearly extrapolate the curves to $V_{x} = 0$, the threshold force
is indistinguishable from zero.
In the inset of Fig.~5 we plot the threshold force $f_c$ extracted from the 
velocity-force curves. 
We have considered
various rates of increasing $f_d$ and find no differences in the 
extracted values of $f_c$.
The threshold decreases with increasing temperature up to $T = 1.5$, 
and above this temperature there is no threshold for motion.  
These results add further evidence
that there is a dynamical freezing or jamming that occurs 
near $T = 1.5$, as also reflected in the defect fluctuations. 
We note that it is beyond the scope of this paper to determine the 
true nature of the finite temperature dynamical freezing, such as whether it
resembles a glass transition. 

\section{Discussion} 

The main goal of this paper is to introduce a new measure, 
the fluctuations in the defect density, for characterizing disordered 
systems near freezing.
The question of whether there is really a finite 
temperature freezing transition
in the equilibrium system near $T = 1.5$ is beyond the scope of our paper. 
Although we
have found several signatures for some form of dynamical slowing 
down at $T = 1.5$, 
it is possible that for very long times beyond our simulation time scales,
the system would show linear diffusion and the fluctuations would lose the 
$1/f$ characteristic at very low frequencies.
We note that there is 
other evidence that topological fluctuations and defects 
appear to be correlated with jamming.  
Recently, the shape features of topological defects 
have been utilized to study the jamming transition
in two-dimensional granular materials
\cite{Abbatecondmat}.  In this case the fluctuations are nonthermal;
however, a jamming transition can still occur. 
Measurements of fluctuations of $P_6(t)$ for the granular system, 
in the manner we suggest in the present work, reveal the same trends that
we have found here.  The defect fluctuations have a white noise
characteristic away from the jamming transition which
crosses over to a $1/f$ signature near jamming with a 
peak in the finite frequency noise power \cite{Abbate}.
This indicates that the topological fluctuations near jamming or 
freezing may universally exhibit 
$1/f$ noise features. 

There has also been recent work on two-dimensional binary glass forming systems
showing correlations between the onset of glassy 
properties and the disappearance
of certain types of topological defects at finite temperature. This adds 
further evidence that there may indeed be a finite temperature freezing or
jamming transition in two-dimensional disordered systems that is connected with
the motion of defects \cite{Hentchel}.          

\section{Summary}
 
To summarize, we have shown that the topological defect fluctuations 
is a useful quantity that can be employed to
understand dynamical freezing and heterogeneities in disordered systems. 
We specifically find that in regimes
where there are dynamical heterogeneities, 
the time dependent  defect density fluctuations show a $1/f$ characteristic 
noise spectrum,
and that at higher temperatures where the motion is uniform, 
a white noise spectrum appears.  The defect noise power also shows
a peak just before the dynamics of the system freezes, and this 
effect is robust for increasing system sizes.
The coordination number lifetimes show a power law distribution 
in the heterogeneous regions which
crosses over to an exponential form at higher temperatures where the 
heterogeneities are lost. 
We correlate the peak in the noise power with the onset of 
subdiffusive behavior 
as well as with the temperature where a threshold force for
the motion of a driven probe particle disappears. 
Our results should be testable for disordered 
charge-stabilized colloidal
assemblies, dusty plasmas, and other disordered charged systems.
We note that recent experiments on granular system near jamming have 
found the same type of behavior in the time series analysis 
of the defect fluctuations
as the jamming transition is approached.  

We thank M.B.~Hastings 
for useful discussions. We also thank A. Abate for sharing his 
unpublished data with us.    
This work was supported by the U.S. DoE under Contract No. W-7405-ENG-36.

\end{document}